\documentclass[a4paper,11pt]{iopart}

\usepackage{color,helvet,mathpazo,epsfig,amssymb}

\usepackage[
pdftitle=Lone pairs in pyrochlores: Ram Seshadri,
pdfstartview=FitH,
bookmarks=false,
colorlinks=true,
citecolor=blue
           ]{hyperref}

\newcommand{\BZN}{(Bi$_{1.5}$Zn$_{0.5}$)(Nb$_{1.5}$Zn$_{0.5}$)O$_7$}
\newcommand{\BT}{Bi$_2$Ti$_2$O$_7$}
\newcommand{\PSn}{Pb$_2$Sn$_2$O$_6$}
\newcommand{\SBT}{SrBi$_2$Ta$_2$O$_9$}

\begin{document}

\title{Lone pairs in insulating pyrochlores: Ice rules and high-$k$ behavior}

\author{Ram Seshadri}

\address{Materials Department and Materials Research Laboratory\\
         University of California, Santa Barbara CA 93106 USA\\
         seshadri@mrl.ucsb.edu}

\begin{abstract}
Pyrochlore dielectric materials such as \BZN\/ (BZN) have
generated interest because they combine high dielectric 
constants with small dielectric loss tangents and yet are cubic at all 
temperatures. The recent low-temperature preparation and structural 
characterization of Bi$_2$Ti$_2$O$_7$, which remains cubic down to 2 K, has 
provided a good model system for understanding the properties of Bi-based 
pyrochlores. In this contribution, the electronic structure of cubic \BT\/ is
visualized and compared with the  
electronic structure of the Aurivillius phase ferroelectric \SBT\/ (SBT), 
which displays a ferroelectric distortion below 608\,K associated with 
the tendency of lone pair active Bi$^{3+}$ to move off-center. Such coherent
off-centering distortions are frustrated on the pyrochlore lattice, and this 
prevents a ferroelectric-paraelectric phase transition in \BT. Instead, 
Bi$^{3+}$ ions in \BT\/ are obliged to off-center in an \textit{incoherent\/} 
manner, that is compatible with the cubic structure being retained. Frustrated
lone pair behavior in the defect pyrochlore \PSn\/ is also described. 
Parallels between the well-studied frustration of certain types of 
\textit{magnetism\/} in pyrochlore compounds (spin-ice) and the striking 
paucity of ferroelectric pyrochlores, arising from the corner-connected 
tetrahedral topology of the pyrochlore lattice are pointed out.

\end{abstract}

\pagebreak

\section{Introduction}

\begin{figure}
\centering \epsfig{file=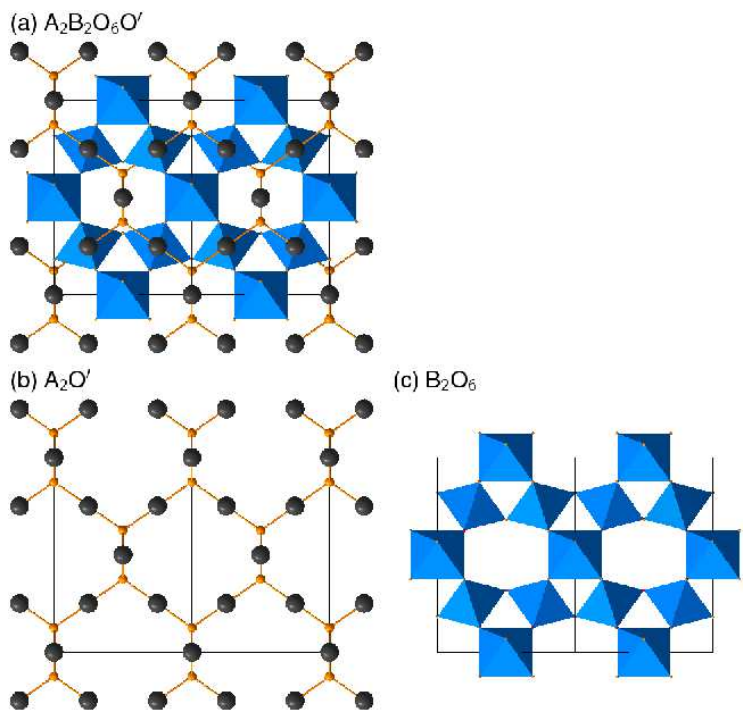, width=10cm}
\caption{(Color) (a) Ideal pyrochlore A$_2$B$_2$O$_6$O$^\prime$ crystal 
structure in the $Fd\bar3m$ space group showing black A atoms, orange 
O$^{\prime}$ and the network of corner-connected BO$_6$ octahedra (blue).
(b) is a view of the tetrahedral A$_2$O$^{\prime}$ sublattice, and (c) is a 
view of the B$_2$O$_6$ sublattice. All views are looking down [110].}
\end{figure}

Pyrochlore oxides\cite{subramanian} have attracted a great deal of recent 
attention as a consequence of their diverse, often unusual, physical 
properties. Recent examples of the diverse properties displayed by pyrochlores
include the discovery of colossal magnetoresistance in Tl$_2$Mo$_2$O$_7$ in 
the absence of multiple valence states,\cite{tl2mn2o7} superconductivity in 
Cd$_2$Re$_2$O$_7$,\cite{hiroi_cd2re2o7} and in the $\beta$-pyrochlore 
KOs$_2$O$_6$,\cite{hiroi_kos2o6} and an unusual chiral magnetic ground state 
in Nd$_2$Mo$_2$O$_7$.\cite{nagaosa}

In the pyrochlore crystal structure, originally associated with the mineral
Ca$_2$Nb$_2$O$_6$F, the B (Nb) atoms form a network of corner-sharing
octahedra. The A-site (Ca) has six O neighbors and two F
neighbors. For the purpose of this work on oxide pyrochlores, the structure
is best described as A$_2$B$_2$O$_6$O$^{\prime}$ comprising two 
interpenetrating sublattices with the formul\ae\/ A$_2$O$^{\prime}$ and BO$_6$.
These views of the structure are displayed in figure.\,1. The crystal chemical 
principles involved in the pyrochlore structure are unusual as has recently 
been emphasized by Vanderah and coworkers.\cite{Vanderah} For example, the
A$_2$O$^{\prime}$ sublattice can tolerate a great deal of displacive disorder,
and consequently, does not obey the usual rules for cation substitution.
These unusual principles play a significant r\^ole in rendering the pyrochlore 
structure suitable for the storage of ions involved in radioactive
waste.\cite{Ewing}

Another crucial feature of the structure is that when regarded separately,
both the A and B sublattices form corner-connected networks of A$_4$ and B$_4$ 
tetrahedra. The formation of such tetrahedral networks can result in
magnetic frustration associated, for example, with the infeasibility 
of decorating the vertices of a tetrahedron with all four spins in an 
anti-parallel configuration.\cite{ramirez_annu_rev,greedan_jmc} 
P. W. Anderson\cite{anderson} first recognized 
the relation between the tetrahedral spin configurations of spinels
and the problem of the crystal structure of ice-$I_h$. According to the 
Bernal-Fowler rules\cite{bernal} each oxygen atom in the ice lattice is 
tetrahedrally bonded to four protons, with the proviso that two of the 
protons are proximal (covalent) and two are distal (H-bonded). 
There is no unique arrangement which satisfies these rules, and as a 
consequence, ice-$I_h$ has a residual zero-temperature
entropy.\cite{pauling,nagle,baxter} 
The crystal structure of ice-$I_h$ is therefore 
frustrated in the sense that it cannot find a perfectly crystalline 
zero-entropy ground state. There is great current interest in such 
frustrated systems, associated with a desire to understand the nature of 
complex potential energy landscapes, and a number of recent studies have
emerged on ``spin-ice'' pyrochlores.\cite{spin-ice,bramwell}
Interestingly, when ground state structures are induced, for example through
OH$^-$ doping, polar ground states become possible in water 
ice.\cite{bramwell_ferroelec}

Unlike ABO$_3$ perovskite oxides where ferroelectricity is not uncommon and is
a well-studied phenomenon,\cite{lines} there exist very few ferroelectric
pyrochlores. The few known system seem to be based on 
Cd$_2$Nb$_2$O$_7$,\cite{cd2nb2o7} and its 
variants.\cite{cdnbo-var1,cdnbo-var2} These
compounds remain the subject of study.\cite{ang_APL_77_732}
``High-$k$'' pyrochlore oxides with large dielectric constants exemplified 
by \BZN\/ (BZN) however do exist and are attracting considerable attention.
BZN is a cubic pyrochlore at all temperatures,\cite{bzn_structure}
with considerable short-range order in the structure.\cite{withers} 
Ceramic samples are known to combine high dielectric constants with low
dielectric losses.\cite{wang,valant} A combination of spectroscopic techniques
reveal that the unusual properties of BZN arise due to local hopping of 
the Bi and O$^\prime$ atoms among several potential minima.\cite{kamba} 
Interestingly, from the viewpoint of applications, these desirable 
characteristics are retained in thin films.\cite{ren,hong,lu1,lu2}

It is intriguing that BZN exhibits all the characteristics of a compound
\textit{about\/} to undergo a phase transition, such as the need to split 
the Bi and O$^\prime$ for satisfactory Rietveld refinements of neutron 
diffraction data,\cite{bzn_structure} and yet remains cubic till the lowest 
temperature. The absence of a phase transition is an important ingredient in 
the usefulness of BZN. For example, in a material with no phase transition, 
there is little fatigue, and the temperature coefficient of the dielectric 
constant is small. 

The preparation of a slightly off-stoichiometric pyrochlore phase related 
to \BT\/ by Sleight and coworkers, \cite{sleight_bi2ti2o7} and the more
recent preparation of stoichiometric, cubic \BT\/ by Hector and 
Wiggin\cite{hector} provide a good model system for understanding 
polar phenomena in the bismuth
pyrochlores. Stoichiometric \BT\/ has been prepared by a low-temperature
route and established by powder neutron diffraction to be a cubic pyrochlore 
down to 2 K, but with extensive site disorder at both the Bi and 
O$^\prime$ sites.\cite{hector} Site disorder on the pyrochlore A-site is 
well-known for lone pair active cations such as Tl$^{+}$,\cite{ganne} 
Sn$^{2+}$,\cite{birchall,cruz} and 
Bi$^{3+}$\cite{bzn_structure,sleight_bi2ti2o7,cava} 
and will be discussed in further detail. 

We have for some time been interested in the stereochemistry of lone pairs in 
extended solids,\cite{seshadri_pmw,seshadri_hill,raulot,seshadri_pias,waghmare}
and have used density functional calculations of electronic structure in 
conjunction with the electron localization function\cite{becke,silvi}
for visualizing lone pair electrons in the real space of crystal 
structures. Here such analysis is extended to the electronic structure of cubic
\BT. For comparison, LMTO calculations have been performed on the Aurivillius 
phase ferroelectric \SBT\cite{smolenskii,subbarao} 
where the transition from a paraelectric to a ferroelectric phase at 608 K is 
driven in large part by the off-centering tendency of the Bi$^{3+}$ ions. 
The electronic structures of paraelectric\cite{stachiotti} and 
ferroelectric \SBT \cite{miura,zhang} have been described previously, but no 
discussion of the disposition of lone pairs in the ferroelectric phase
has been presented. The disposition of Pb$^{2+}$ lone pairs 
in the cubic defect pyrochlore \PSn\cite{morgenstern} are also examined.

\section{Details of computation}

Linear Muffin-Tin Orbital calculations\cite{andersen1,andersen2} 
on pyrochlore \BT\/
were performed within the atomic sphere approximation using version 47C of 
the Stuttgart \textsc{tb-lmto-asa} program.\cite{stuttgartLMTO} 
Scalar-relativistic Kohn-Sham equations within the 
local density approximation \cite{vonBarth} were solved taking all relativistic
effects into account except for the spin-orbit coupling. The calculations
were performed on 72 irreducible $k$ points within the primitive wedge of 
the Brillouin zone. LMTO electronic structures were analyzed by 
calculating crystal orbital Hamiltonian populations (COHPs) and electron 
localization functions (ELFs). The crystal orbital Hamiltonian population 
COHP\cite{cohp1,cohp2} is a tool which permits energy-resolved analysis of 
bonding, The ELF provides a measure of the local influence of the Pauli 
repulsion on the behavior of electrons and permits the mapping in real space 
of core, bonding, and non-bonding regions in a crystal.\cite{becke,silvi}
Calculations were performed on the ideal crystal structure of pyrochlore
Bi$_2$Ti$_2$O$_7$\cite{hector} (space group $Fd\bar3m$, 
$a$ = 10.37949\,\AA), displayed in figure\,1. The atom positions are Bi (0,0,0),
Ti ($\frac 1 2$,$\frac 1 2$,$\frac 1 2$), O ($\frac 1 8$,$\frac 1 8$,$x$)
and  O$^\prime$ ($\frac 1 8$,$\frac 1 8$,$\frac 1 8$). $x$ = 0.43128 was taken
from the reported room temperature crystal structure refinement.\cite{hector}
LMTO calculations were also performed on the $A2_1 am$ crystal structure 
of ferroelectric \SBT\/ taken from the detailed structural study
of Rae \textit{et al.}\cite{rae} A grid of 72 irreducible $k$ points within 
the primitive wedge of the Brillouin zone was used for the calculation on
\SBT. The crystal structure of the defect pyrochlore \PSn\cite{morgenstern}
is reported in the International Crystal Structure Database (entry 15308)
to be a cubic pyrochlore with $a$ = 10.719\,\AA. Atoms are Pb (0,0,0), 
Sn ($\frac 1 2$,$\frac 1 2$,$\frac 1 2$), and O ($\frac 1 8$,$\frac 1 8$,$x$) 
with $x$ = 0.4250. O$^\prime$ is missing in this structure. LMTO calculations
on \PSn\/ employed a grid of 72 irreducible $k$ points.

\section{Results}

\subsection{\BT}

\begin{figure}
\centering \epsfig{file=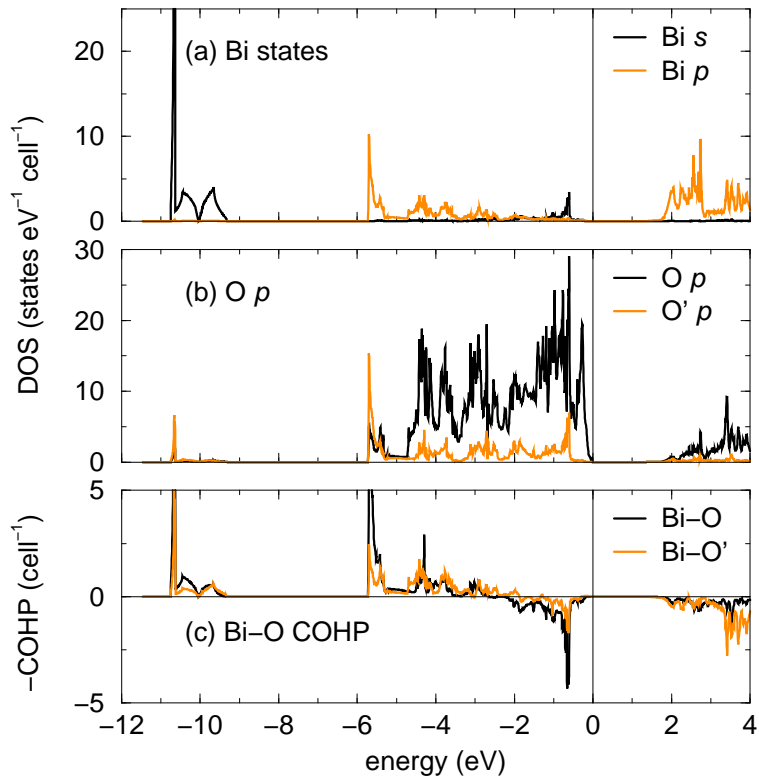, width=10cm}
\caption{(Color) 
(a) Densities of Bi $s$ and $p$ states for the ideal cubic structure
of \BT. (b) Densities of the $p$ states of O and O$^\prime$ in \BT. COHPs 
showing all Bi-O and Bi-O$^\prime$ interactions within the primitive unit 
cell of \BT. The origin on the energy axis is the top of the valence band.}
\end{figure}

\begin{figure}
\centering \epsfig{file=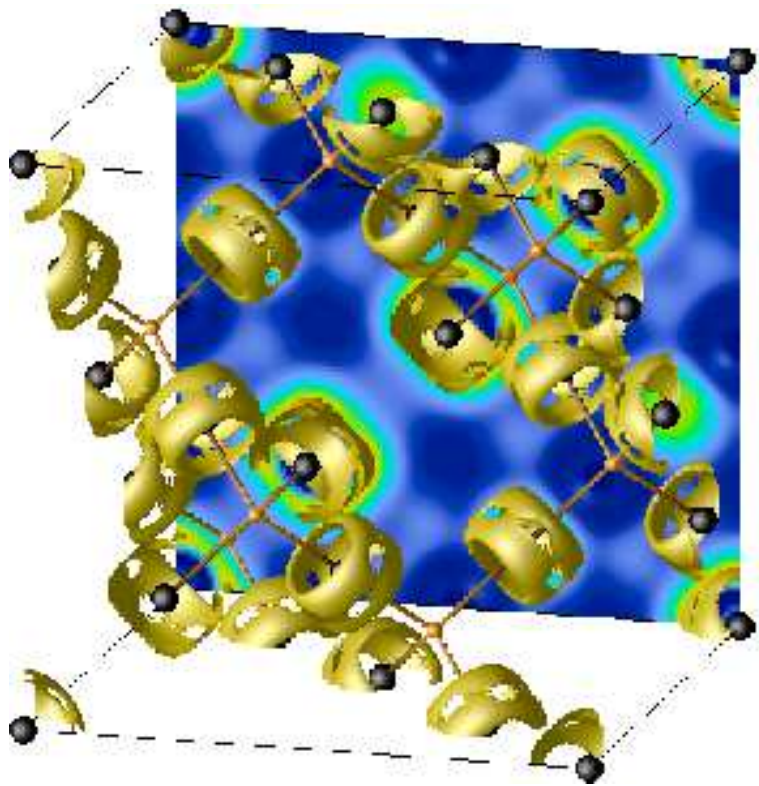, width=10cm}
\caption{(Color) Bi-O$^\prime$ network of \BT\/ showing the valence electron 
localization isosurfaces visualized for a value of ELF = 0.65. 
The total ELF for a single plane is projected on the back of the unit cell.
The localization scale runs from deep blue (ELF = 0) to white (ELF = 1).} 
\end{figure}

The Bi$_2$O$^\prime$ network of \BT\/ can be imagined as being 
formed of a diamond lattice of O$^\prime$, with one Bi atom at the center of
every O$^\prime$-O$^\prime$ linkage. The Bi$_2$O$^\prime$ network thus
satisfies one of the Bernal-Fowler rules\cite{bernal}
in that each O$^\prime$ is tetrahedrally surrounded by Bi. The network
does not satisfy the ``two-near and two-far'' rule, meaning that ideal 
pyrochlore \BT\/ is perfectly crystalline and has no residual entropy.

The projected Bi and oxygen LMTO densities of state (DOS), and the Bi-oxygen
COHPs of cubic \BT\/ are displayed in the three panels of figure\,2.   
The band gap in cubic \BT\/ is about 2 eV. Bi $s$ states in figure~2(a) 
are seen to reside in a very narrow window of energy around -10 eV with respect
to the top of the valence band (set as the origin of the energy axis in this 
and other plots). Bi $p$ 
states, as expected for a formally Bi$^{3+}$ system, are largely empty and 
form the conduction band which starts at 2 eV. Some Bi $p$ states are found
mixed with the valence band as a result of covalency with oxygen. Bi $s$ and 
$p$ states of \BT\/ are resemble those previously observed 
in cubic ``pre-distorted'' BiMnO$_3$.\cite{seshadri_hill}
O $p$ and O$^\prime$ $p$ states are displayed in figure\,2(b). There are 
six O for every one O$^\prime$, giving rise to a large difference in
the relative number of states. O$^\prime$ states are particularly 
prominent around -5.5 eV, near the bottom of the valence band, where O$^\prime$
$p$ forms band with Bi $p$ states. Bi-O and Bi-O$^\prime$ COHPs are 
displayed per primitive unit cell (or for a single BiO$_6$O$^\prime _2$ 
polyhedron) in figure\,2(c). The integrated COHP which is indicative of the 
strength of interaction, confirms that Bi $s$ states are quite inert in the 
cubic crystal structure. Stronger Bi $p$ and O$^\prime$ $p$
interactions manifest as a large bonding COHP at the bottom of the valence 
band. Because of the longer distance between Bi and O (2.635\,\AA) compared 
with Bi and O$^\prime$ (2.247\,\AA), the Bi-O interaction is weak and O obtain 
most of their bond valence from Ti.

The valence electron localization function (ELF) of \BT\/ is plotted within
the real space of the crystal structure in figure\,3. For clarity, Ti and O 
atoms are not displayed. Bi and O$^\prime$ are connected to emphasize the 
``interrupted'' diamond lattice. Isosurfaces of the ELF for a value of 0.65 
manifest the lobes of $s$ electrons around Bi. These lobes are constrained by 
the symmetry of the structure to form cylindrical objects with the cylindrical 
axis along different [111] directions, corresponding to the Bi-O$^\prime$ 
bonds. The ELF is also displayed on the back plane of the crystal structure as 
a map running from deep blue (poorly localized regions) to white (strongly 
localized regions). The particular manner in which the lone pairs localize 
around Bi atoms is an artifact of the ideal crystal structure is used here. 
In reality, both Bi and O$^\prime$ are shifted from their crystallographic 
sites,\cite{hector} albeit in a incoherent manner. To understand what the lone 
pair localization should more properly resemble, the disposition of lone pairs 
in the Aurivillius phase ferroelectric \SBT\/ is examined in the following
subsection.  

\subsection{\SBT} 

\begin{figure}
\centering \epsfig{file=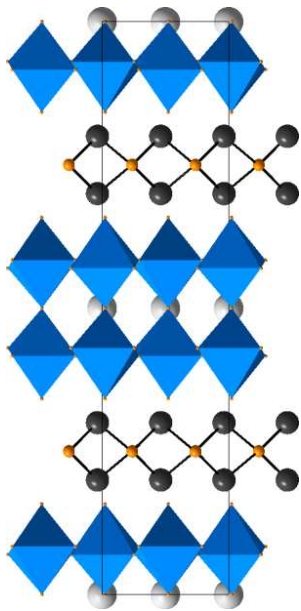, width=4cm}
\caption{(Color) Crystal structure Aurivillius phase SrBi$_2$Ta$_2$O$_9$
projected down the $b$ axis. The black spheres are Bi, orange are O, grey are
Sr and blue octahedra surround Ta.}
\end{figure}

\begin{figure}
\centering \epsfig{file=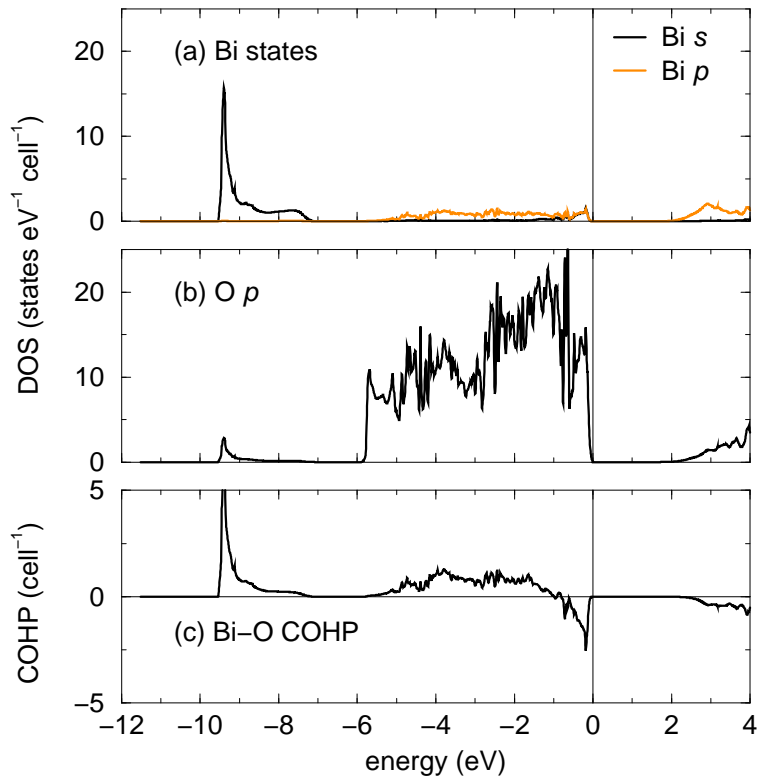, width=10cm}
\caption{(Color) (a) Densities of Bi $s$ and $p$ states in \SBT.
(b) Densities of the all $p$ states of O \SBT. COHPs 
showing all Bi-O interactions within the primitive unit 
cell of \SBT. The origin on the energy axis is the top of the valence band.}
\end{figure}

\begin{figure}
\centering \epsfig{file=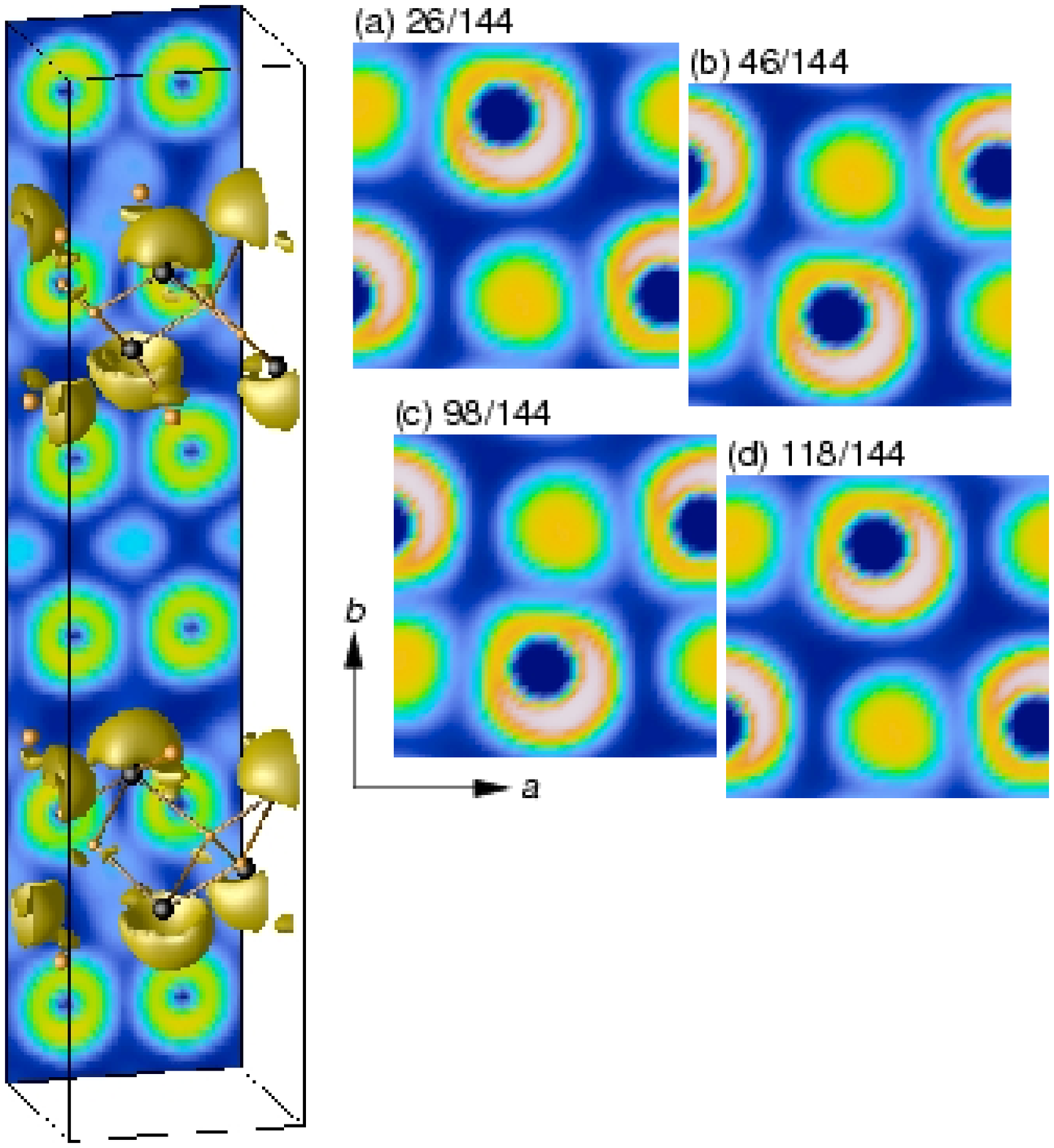, width=10cm}
\caption{(Color) ELF isosurface projected within the unit cell of \SBT\/ for
an ELF value of 0.65. The panels (a-d) on the right are planes of the ELF
parallel to the $c$ axis; the planes are chosen to contain Bi atoms. 
The values of $z/c$ of the different planes are indicated as ratios.}
\end{figure}

The  $A2_1 am$ crystal structure of \SBT\cite{rae}
is displayed in figure\,4 and comprises alternate stacks of double-perovskite
[SrTa$_2$O$_7$]$^{2-}$ slabs interleaved with [Bi$_2$O$_2$]$^{2+}$ slabs. 
The latter are reminiscent of the Bi-O sublattice of BiOF\cite{seshadri_pias}
where Bi lone pairs organize themselves in an antiferrodistortive manner, 
akin to what is seen in PbO.\cite{raulot,seshadri_pias} In this 
\textit{distorted\/} crystal structure, Bi $s$ states [figure\,5(a)] are 
significantly more disperse than was observed for cubic \BT. 
Relatively few Bi $p$ states are observed in the region of filled O $p$ states 
[figure\,5(b)]; most Bi $p$ states are empty and, starting at 2 eV, form the
conduction band along with empty Ta $d$ states (not displayed). The Bi-O COHP
displays quite strong bonding in the region of Bi $s$ states. Both Bi $s$ 
states and O $p$ states are filled, so their interaction gives rise to filled
bonding states around -9 eV and filled antibonding states just below the top
of the valence band. The intermediation of empty Bi $p$ states
provides net stabilization.\cite{waghmare} The Bi-O COHP of \SBT\/ displayed 
in figure\,5(c) is what the Bi-O COHP of \BT\/ displayed in figure~2(a) would 
resemble were it permitted to take on an appropriately distorted 
coordination.

The valence ELF of \SBT\/ is displayed in the different panels of figure\,6.
The left panel of figure\,6 displays the ELF isosurface 
within the space of the unit cell. For clarity, other atoms than in the Bi-O 
network have been let out of the depiction.
The net ``polarization'' of the lone pair lobes is more 
clearly seen in projections down the $c$ axis [figure\,6(a-d)]. 
The ELF was been calculated on a 48$\times$48$\times$144 ($a \times b 
\times c$) real-space grid within the unit cell. Figure\,6(a-d) shows four 
sections of the ELF at four different heights along the $c$ direction, 
corresponding to positions of Bi atoms in the unit cell. In this view, it is 
clear that all the Bi lone pairs have a component that points in the $a$ 
direction, meaning that the Bi are shifted slightly along
$b$. It is this in-plane shift which gives \SBT\/ its polarization. The very 
different shapes of the lone pair lobes associated with centered Bi in
\BT\/ (figure\,3) and with off-centered Bi in \SBT\/ (figure\,6) should be 
noted. 

\subsection{\PSn}

\begin{figure}
\centering \epsfig{file=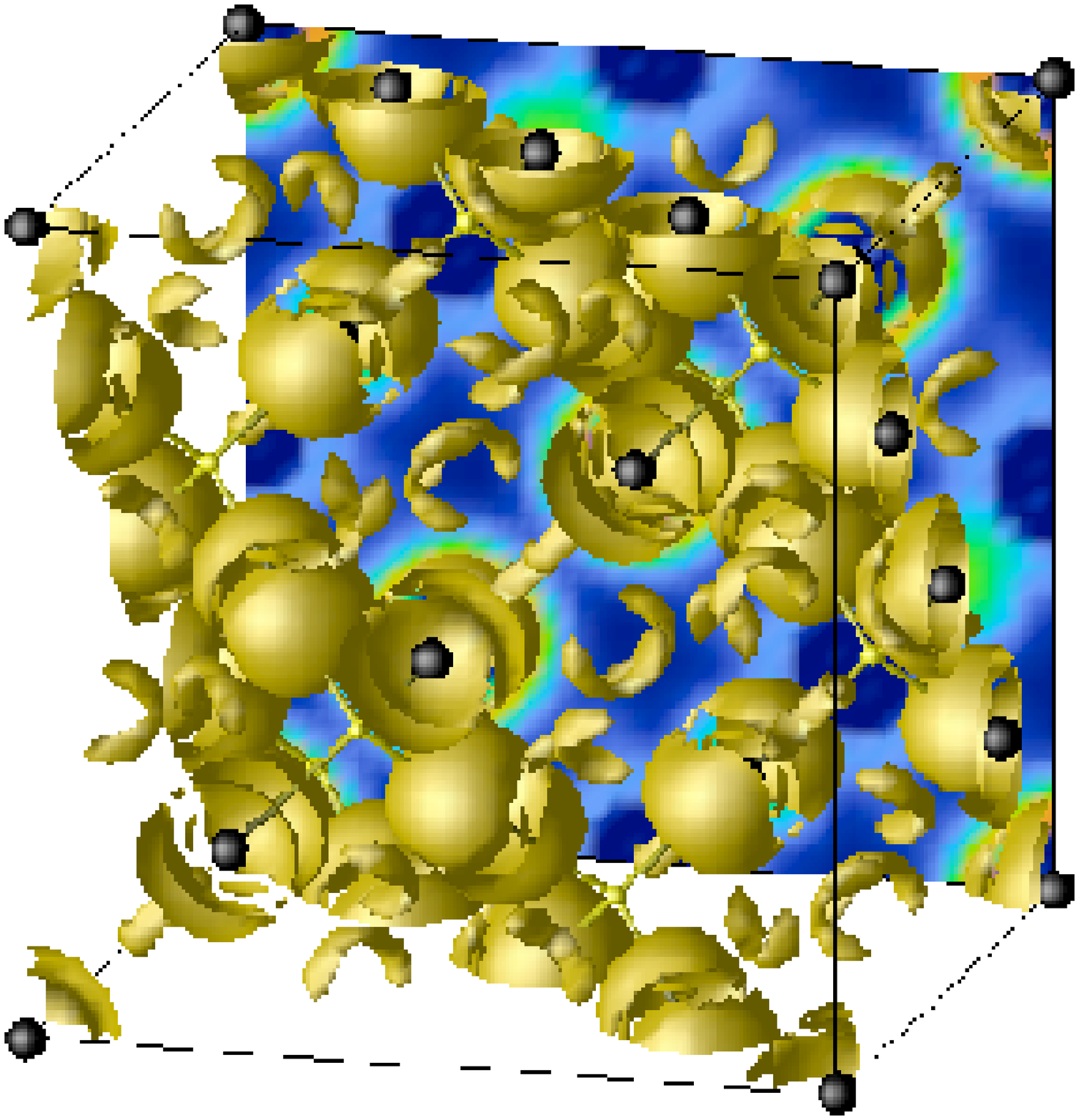, width=10cm}
\caption{(Color) ELF isosurface projected within the unit cell of \PSn\/ for
an ELF value of 0.65. The yellow spheres are vacant positions corresponding
to O$^\prime$ in the ideal pyrochlore.}
\end{figure}

Morgenstern-Badarau and Michel\cite{morgenstern} have reported the defect 
pyrochlore
crystal structure of ``Pb$_2$Sn$_2$O$_6 \cdot x$(H$_2$O)''. Interestingly, the 
compound has long history and could be one of the components of a yellow 
pigment used by renaissance painters 
called lead-tin yellow.\cite{Harris} The O$^\prime$ atom is missing in this 
crystal structure and is replaced by a vacancy ($\Box$). A$_2$O$^\prime$ 
in the usual pyrochlore then corresponds to Pb$_2$$\Box$. It is natural to 
assume that Pb$^{2+}$ lone pairs would occupy the site that is normally taken 
up by O$^\prime$. This would parallel the crystal-chemical relation between 
litharge PbO and PbO$_2$; the former having lone pairs in the place of the 
oxygen atoms of the latter.\cite{hyde} However the A$_2$O$^\prime$ network of 
the pyrochlore lattice depicted in figure\,1 allows us to recognize that for 
every four Pb atoms, there are two sites into which the lone pairs can 
localize.  The problem now maps precisely on to the 
Bernal-Fowler rules with the lone pair-Pb connection corresponding to the 
short (covalent) O-H interaction. The valence ELF of cubic \PSn\/ displayed 
in figure\,7 support this argument. The symmetry of the crystal structure 
prevents localization into a single lobe, and instead, a bi-lobed lone pair 
is obtained. Topological frustration again prevents coherent ordering of the 
lone pairs in some distorted low-temperature structure. Re-examination
of the crystal structure of defect pyrochlore \PSn\/ should reveal 
disordering on this site; splitting every Pb atom into two along the
$\Box$-Pb-$\Box$ direction (the [111] directions). The precise nature of such 
splitting would therefore be distinct from what is observed for Bi in \BT\/
where it is forms an annulus around O$^\prime$-Bi-O$^\prime$.\cite{hector} 

\section{Implications}

\begin{figure}
\centering \epsfig{file=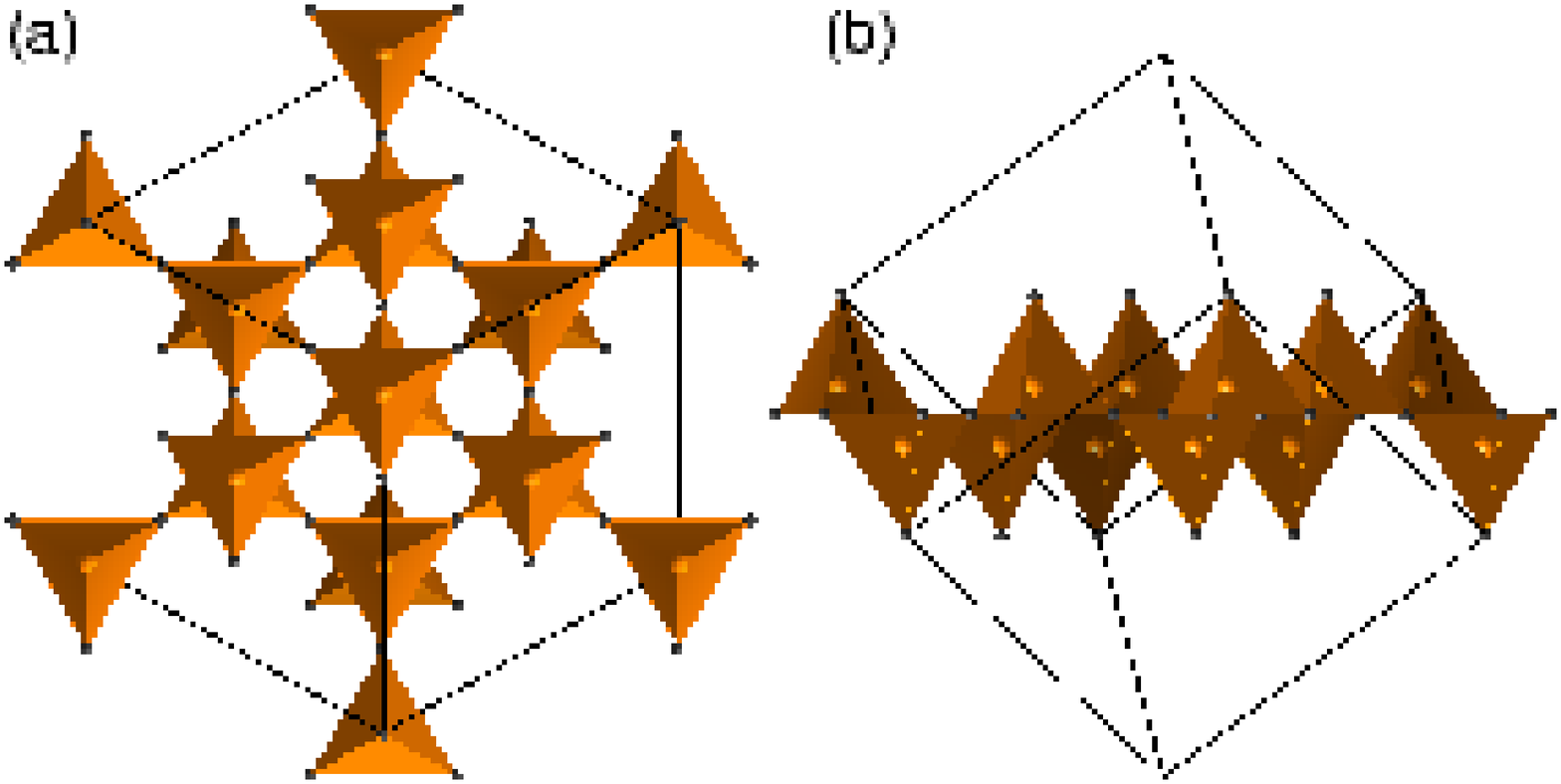, width=14cm}
\caption{(Color) (a) Corner-connected O$^\prime$A$_4$ tetrahedra of the 
A$_2$B$_2$O$_6$O$^\prime$ structure projected down [111] (b) A single slab
of the network seen from the side; the pyrochlore structure described
as a stellated \textit{kagom\'e\/} network.}
\end{figure}

\begin{figure}
\centering \epsfig{file=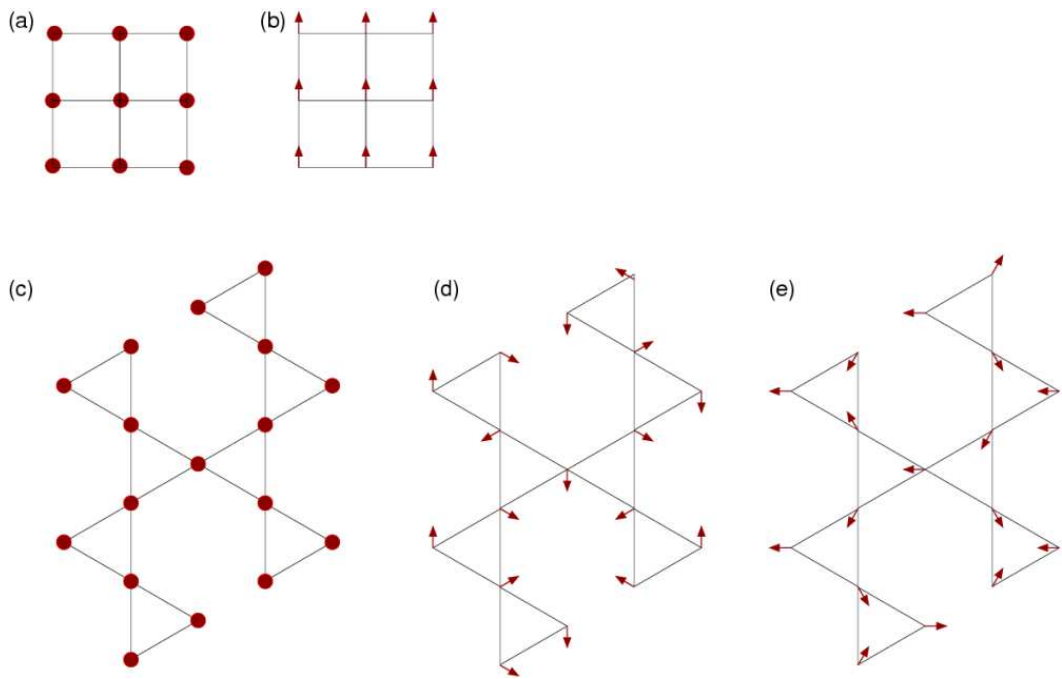, width=14cm}
\caption{(Color) (a) is a square lattice with lone-pair active ions in the
corners. Coherent localization of the lone pairs results in a polar
rectangular lattice (b). (c) is a \textit{kagom\'e\/} 
lattice with lone pair active
ions at the corners. The lone pairs can either localize towards other vertices
(d) or towards the centers of the triangle (e), but they do not do so 
coherently.}
\end{figure}

To aid the rest of the discussion, the panels of figure\,8 display
two different views of the O$^\prime$-centered tetrahedral network, also 
referred to as the tridimyte sublattice,\cite{Vanderah} of the pyrochlore
structure. The two views suggest that we could view the pyrochlore structure
as a stellation followed by stacking of \textit{kagom\'e\/} lattices of A 
atoms.  A \textit{kagom\'e\/} lattice is a lattice of corner sharing 
triangles. The stellation of this lattice is achieved by making each triangle 
a tetrahedron by alternately capping the triangles above and below the planes 
they are in and creating tetrahedra pointing either up or down (along [111] 
in the pyrochlore structure). In 2D the pyrochlore lattice should be 
well-represented by a \textit{kagom\'e\/} lattice. This allows us to compare 
in a simple manner, the topologies of perovskites, which we represent by a 
square lattice, and the pyrochlore, which we represent using the 
\textit{kagom\'e\/} lattice (figure\,9).

A square lattice is easily distorted to a rectangular lattice [figure\,9(a) 
and (b)] allowing \textit{coherent\/} localization of lone pair lobes, which
we indicate with arrows. This simple description is reasonably accurate in 
describing the pyroelectric to ferroelectric (cubic $Pm\bar 3m$ to tetragonal 
$P4mm$) phase transition in perovskite PbTiO$_3$.\cite{lines,seshadri_pias} 
 
A \textit{kagom\'e\/} 
lattice with lone pair-active ions at its corners [figure\,9(a)]
could localize lone pairs in two different ways: The first is for the 
lone pairs to localize perpendicular to the line joining the centers of the 
triangles, as depicted in figure\,9(c). This is the case with the annular 
lone-pair localization in \BT. The second is for the lone pairs to localize 
towards the centers of triangles, as depicted in figure\,9(d). This is 
representative of \textit{kagom\'e\/} spin ice\cite{kagome-spin-ice} and is 
similar to what is observed \PSn. In neither case is it a simple matter
to arrange the arrows such that a unit cell is obtained. Note that neither
of these arrangements would be \textit{ferrodistortive}. In fact, while we
have focused on the paucity of ferroelectric pyrochlores, our analysis
actually refers to any symmetry lowering phase transition.

In the rare instances when pyrochlore oxides distort as a result of coherent 
ordering of stereochemically active lone pairs, the resulting crystal 
structures can be be quite complex, for example, as determined recently for 
Bi$_2$Sn$_2$O$_7$.\cite{evans} However, retention of the cubic structure is 
important for properties. For instance, the pyrochlore composition 
Bi$_2$Zn$_{1/3}$Nb$_{4/3}$O$_7$ crystallizes in a monoclinic structure related 
to zirconolite,\cite{levin} and the material is not as interesting as \BZN\/ 
for high-$k$ applications. Wang \textit{et al.}\cite{wang} have presented a 
simple and powerful argument for when A$_2$B$_2$O$_7$ pyrochlores with lone 
pair active A cations distort. They point to a relative competition between 
covalency in the A-O and B-O networks. Pyrochlores with strongly covalent 
B-O networks tend to retain the cubic structures. 

The arguments presented here bear close similarities to the frustration of soft modes in ZrW$_2$O$_8$, proposed by Ramirez and 
coworkers\cite{Ramirez_zrw2o8} as being responsible for its negative thermal 
expansion behavior.\cite{mary} Recently, Loidl
and coworkers\cite{loidl} reported unusual magnetoferroelectric properties
of spinel CdCr$_2$S$_4$. In this compound, relaxor ferroelectricity is observed
in the absence of any substitutional disorder. The authors suggest that the
origin of such relaxor behavior could be the topological frustration of a true 
ferroelectric ground state. In the superconducting $\beta$-pyrochlore oxides, 
Kune\v{s}, Jeong, and Pickett\cite{kunes} have used first principles 
calculations to determine that local displacements of the A cation are 
critical in determining superconducting $T_c$s. One could argue that in the 
absence of the topological frustration provided by the pyrochlore lattice, 
these local displacements would lead to a distorted ground state structure. 
In a related vein, for macroscopic materials built up from bars and pins, the 
unusual properties of \textit{kagom\'e\/} lattices with respect to the 
applications of stresses have already been noted.\cite{Guest}

To summarize, it is proposed that in pyrochlores such as \BZN, properties are 
dominated 
by topological frustration rather than cation site disorder. \BT\/ itself 
is therefore a candidate for high-$k$ behavior, and should be regarded as 
a parallel of systems such as cubic, \textit{predistorted\/} PbTiO$_3$. 
In relaxor ferroelectrics, substitutional disorder frustrates long-range 
ordering and gives rise to unusual behavior, such as broad dielectric 
anomalies as a function of temperature.\cite{relaxor1,relaxor2} 
If relaxor ferroelectrics with site disorder are the ferroelectric equivalent 
of substitutional spin glasses such as Cu:Mn, then pyrochlore dielectrics could 
be considered the ferroelectric analogues of topologically frustrated magnets. 
In analogy with topologically frustrated magnets, the recording of  
low-temperature heat capacity of \BT\/ 
should permit the residual entropy to be measured and compared with values 
found in pyrochlore spin-ice.\cite{spin-ice} Finally, in analogy with spin-ice,
it is proposed that the term ``charge-ice'' is used to describe frustrated
pyroelectric phenomena in systems such as spinels and pyrochlores.
Topological frustration in polar materials could constitute a powerful design 
parameter for useful and novel materials properties.

\section{Acknowledgements}
I thank Susanne Stemmer for bringing \BZN\/ to my attention, and for 
valuable discussions and references. Tony Evans, Alois Loidl, Bob McMeeking, 
Art Ramirez, Ivana Radosavljevic-Evans, Nicola Spaldin, Roser Valent\'{\i}, 
Terrell Vanderah, and Pat Woodward are also thanked for discussions. Support 
from the National 
Science Foundation through the MRSEC program under award no. DMR00-80034, and 
through a Chemical Bonding Center (Chemical Design of Materials) under award 
no. CHE04-34567 is gratefully acknowledged.

\end{document}